\documentclass{aa}

\usepackage{graphics}

\begin{document}

\newcommand{\gtrsim}{~\raise.5ex\hbox{$>$}\kern-.8em\lower 1mm\hbox{$\sim$}~}
\newcommand{\lesssim}{~\raise.5ex\hbox{$<$}\kern-.8em\lower 1mm\hbox{$\sim$}~}

\thesaurus{3(11.17.3; 12.12.1; 13.25.3; 01.13.2; 12.04.2)}

\title{The large-scale diffuse X-ray emission surrounding quasars: an 
investigation using the scaling-index method}

\author{K. A. Williams\thanks{\emph{Present address:} UCO/Lick Observatory and
Board of Astronomy and Astrophysics, University of California, Santa Cruz, CA
95064, USA} \and W. Brinkmann \and G. Wiedenmann}

\offprints{K. Williams (williams@ucolick.org)}

\institute{Max--Planck--Institut f\"ur extraterrestrische Physik,
 Giessenbachstrasse, D-85740 Garching, FRG}

\date{Received ? / Accepted ?}

\titlerunning{Diffuse X-ray emission surrounding quasars}
\authorrunning{K.A. Williams et al.}

\maketitle

\begin{abstract}

The large-scale ($\sim 20\arcmin$) diffuse x-ray background surrounding a
sample of quasars observed during \emph{ROSAT} PSPC pointed observations
is studied using a new source-detection algorithm, the scaling-index method.
This algorithm, which can identify individual photons as belonging to a 
source or background, is useful for detecting faint, extended sources in
noisy fields. Using this method, we find that, contrary to conclusions 
drawn by others, there is scant evidence for preferential enhancement of
x-ray backgrounds surrounding radio-loud quasars by foreground x-ray
emitting clusters of galaxies. Rather, all fluctuations in these backgrounds
can be explained by varying levels of emission from a galactic thermal
plasma of temperature $T\approx 0.14$ keV. No difference is observed
 between the diffuse x-ray backgrounds of radio-loud and radio-quiet quasars.

\keywords{quasars: general -- large-scale structure of Universe -- X-rays: general -- Methods: data analysis -- diffuse radiation}
\end{abstract}

\section{Introduction}
\smallskip

It is generally agreed that radio-loud quasars and radio-quiet quasars
reside in markedly different environments for redshifts $z \gtrsim 0.6$, with 
radio-loud quasars lying in rich clusters and radio-quiet quasars residing
in smaller clusters (Boyle \cite{boyle}), 
although some evidence has been introduced to the contrary 
(Fried \cite{fried}).  Burg et al.~(\cite{burg}) and Briel \& Henry 
(\cite{briel}) have shown that rich clusters, namely, Abell clusters, are 
luminous x-ray sources and that the cluster richness is correlated with the
 x-ray  luminosity. Na\"{\i}vely combining these two results 
leads to the conclusion that the background x-ray emission surrounding 
radio-loud quasars should be higher than the background emission surrounding 
radio-quiet quasars.

Unfortunately, the expected evolution of clusters with time  does not permit 
us to make such a straightforward conclusion.  Hall et al. ~(\cite{hall}) 
summarize arguments
where, depending upon the model used to explain the
associations of radio-loud quasars with richer host clusters, the
x-ray luminosity of the host cluster may be weaker than expected, especially 
if the host clusters are not virialized.  The 
search of Hall et al. for x-ray emission from the host clusters of
two moderate-redshift quasars failed to find any such emission.  This result 
suggests that the direct detection of x-ray emission from quasar host clusters
would be very difficult.

Bartelmann et al.~(\cite{bartelmann}, hereafter BSH) searched for diffuse 
x-ray emission 
around radio-loud quasars using \emph{ROSAT All-Sky Survey} (RASS) data.  
They claim
to have found significant detections of diffuse emission on angular scales
of $\gtrsim 10'$ for distant ($z \gtrsim 1.5$) and nearby ($z \lesssim 1.0$) 
radio-loud quasars, while radio-loud quasars with intermediate distances 
displayed no significant excess diffuse emission.  They argue that the 
sources of the excess emission surrounding the distant quasars are most 
likely foreground clusters of galaxies, themselves diffuse x-ray emitters.
They estimate an excess count rate on the order of $10^{-2}$ s$^{-1}$ in 
the RASS due to these hypothetical clusters.

In order to study further the correlation of quasars with x-ray emission and
in order to compare the diffuse x-ray backgrounds surrounding radio-loud and
radio-quiet quasars, we undertook a comparison study of the x-ray backgrounds
surrounding quasars observed with the \emph{ROSAT} 
Position-Sensitive Proportional Counter (PSPC) (Pfeffermann et al. 
\cite{pfeffer}) during the pointed-observation mode of the mission.  The
main advantage of the pointed observations over the RASS
observations is a large increase in sensitivity.  The typical observing time 
for RASS fields near the ecliptic plane is 400 sec 
(Voges et al. \thinspace\cite{voges}), while the pointed observations 
discussed here had net exposure times of $\ge10$ ksec. Such an 
observing time also should permit us to detect the hypothetical clusters of
BSH. The pointed observations also have 
a higher angular resolution, which aids in source detection.
One disadvantage to the use of pointed observations is the 
selection effect introduced by the pointed observations themselves; 
presumably, mainly ``interesting'' objects were targeted by the initial
observer. This introduces unknown and unquantifiable biases into
our quasar sample. 

Additionally, the work of BSH compared backgrounds
surrounding radio-loud quasars to ``blank'' fields. In this study,
we compared the radio-loud quasar backgrounds to radio-quiet quasar
backgrounds rather than blank fields. According to McHardy  et al. 
(\cite{mchardy}), deep \emph{ROSAT} images result in a surface density of QSOs
of $129\pm 28$ deg$^2$ for R-band magnitudes brighter than 21. For our
fields, this would imply around 45 QSOs in each field. Even if only a small
fraction of these were detectable in a 10 ksec exposure, the search for
control fields devoid of QSOs would be formidable. 

\section{Data Analysis}
\smallskip
\subsection{Sample Selection}
The sample consists of quasars from the catalogue of 
V\'eron-Cetty \& V\'eron (\cite{veron}) viewed during \emph{ROSAT} PSPC
pointed observations, either as targeted or serendipitously-observed objects.
All selected observations are available in the public \emph{ROSAT} data 
archive. This list contains 127 quasars, of which 27 have been detected in 
the radio band.

In order to obtain a uniform sample, we selected 10\,050 seconds of observing
time from each PSPC observation in twenty-five 402-second intervals, 
each interval
corresponding to the wobble period of the \emph{ROSAT} spacecraft. The particle
background was minimized by selecting only those intervals where the Master
Veto Rate was between 40 and 170 counts s$^{-1}$ and the oxygen column density 
was less than $1.0\times 10^{15}$ cm$^{-2}$. 
Further contamination was removed by examining the accepted and transmitted
count rate and rejecting time intervals where this rate was markedly above
the average, which varies depending on the total flux of the targeted
field. We minimized the contamination of the extragalactic background from 
galactic emission by selecting only those events with an 
amplitude $> 60$, corresponding to photon energies $\gtrsim 0.6$ keV. We
rejected fields where the candidate quasar was located outside the central 
$20\arcmin$ of the PSPC field of view in order to minimize vignetting and 
off-axis image degradation.  These selection criteria resulted in the
reduction of the sample to 72 quasars.

The sample size was further reduced by rejecting fields where the 
quasar was not detected by the scaling index method (Wiedenmann et al. 
\cite{wiedenmann}, hereafter WSV, see \S 2.2).  We rejected four 
additional 
fields, containing a total of five quasars, due to their location in areas of
extended galactic hard x-ray emission such as Loop I.  One quasar near the
x-ray bright galaxy cluster \object{Abell 1795} was rejected due to the 
large angular
extent of the cluster, which introduced very large uncertainties in the
background corrections.  Two further fields containing one sample quasar
each were rejected due to anomalous light curves of the observed background
(variations greater than $2\sigma$ in the observed count rate).  

The remaining 33 quasars constituting the sample used in our analyses 
are listed in Tab.~\ref{tab:1}.  The table includes the names and redshifts
of the objects as listed in V\'eron-Cetty \& V\'eron.  The listed coordinates 
are precessed to J2000.0 from the coordinates in the catalogue. The 
start-of-observation dates for each image (in \emph{ROSAT} days) is provided,
as is the adopted neutral hydrogen value for each quasar (see \S 2.3). 
Columns then list the total number of background photons in the accepted fields
as measured from both the maximum-likelihood and scaling-index method
algorithms described below.  Next, the detected photon fluxes in photons 
sec$^{-1} \deg^{-2}$ are 
listed. We did not attempt to unfold the events through the PSPC detector
response matrix.

\begin{table*}
\caption{Quasar sample after defined selection process.  The horizontal line
separates radio-quiet QSOs from radio-loud QSOs (above and below line, 
respectively)}
\label{tab:1}
\begin{tabular*}{17.5 cm}{l@{\extracolsep{\fill}}rrcccccc}\hline\hline
Name & RA ($J2000$) & Dec ($J2000$) & z & Date of Obs.$^{(a)}$ & ${N_H}$ 
(${10^{20}}$ cm$^{-2}$) &
 $ {N_{ML}}^{(b)} $ & ${N_{SIM}}^{(b)}$ & ${\Phi_{SIM}}^{(c)}$ 
 \\ \hline
 \object{E 0015+162} & 0 18 31.9 & 16 29 26 & 0.554 & 782.2 & 4.07 & 2701.6 & 2762.1 & 0.7470 \\
 \object{PHL 6625} & 0 46 51.8 & $-$20 43 30 & 0.380 & 576.1 & 1.57 & 2511.1 & 2652.6 & 0.7174\\
 \object{SGP 2:20} & 0 51 52.9 & $-$29 15 00 & 0.601 & 742.3 & 1.80 & 2531.8 & 2685.1 & 0.7262 \\
 \object{SGP 3:19} & 0 54 59.1 & $-$28 14 30 & 0.779 & 590.4 & 1.80 & 2531.6 & 2717.0 & 0.7348 \\
 \object{SGP 3:39} & 0 55 43.3 & $-$28 24 09 & 1.964 & 590.4 & 1.89 & 2540.4 & 2724.5 & 0.7368 \\
 \object{SGP 4:41} & 0 57 24.5 & $-$27 31 60 & 1.209 & 774.2 & 1.86 & 2142.2 & 2375.6 & 0.6425 \\
 \object{E 0121+034} & 1 24 33.2 & 3 43 35 & 0.336 & 788.0 & 3.37 & 2503.0 & 2555.0 & 0.6910 \\
 \object{NGC 520.40} & 1 24 57.5 & 3 53 48 & 1.205 & 788.0 & 3.27 & 2493.4 & 2547.4 & 0.6889 \\
 \object{Q 0123-005B} & 1 26 02.2 & $-$0 19 24 & 1.761 & 788.0 & 3.36 & 2057.2 & 2263.8 & 0.6122 \\
 \object{MS 02074-1016} & 2 09 56.8 & $-$10 02 51 & 1.970 & 602.1 & 2.22 & 2329.3 & 2453.1 & 0.6634 \\
 \object{QSF 3:31} & 3 41 55.8 & $-$44 16 37 & 1.797 & 1177.2 & 1.66 & 2065.8 & 2295.2 & 0.6207 \\
 \object{PG 1115+080} & 11 18 17.0 & 7 45 60 & 1.722 & 550.7 & 3.62 & 2555.8 & 2578.9 & 0.6974 \\
 \object{PG 1116+215} & 11 19 08.7 & 21 19 18 & 0.177 & 367.1 & 1.40 & 2707.7 & 3003.2 & 0.8121\\
 \object{US 2694} & 11 36 55.0 & 29 51 32 & 1.858 & 368.1 & 1.77 & 2417.6 & 2735.6 & 0.7398\\
 \object{PG 1411+442} & 14 13 48.3 & 44 00 14 & 0.089 & 396.2 & 1.05 & 2700.6 & 2874.9 & 0.7775\\
 \object{QS M5:42} & 22 02 29.9 & $-$19 01 52 & 1.045 & 555.9 & 2.87 & 3042.8 & 3123.4 & 0.8447 \\
 \object{MS 22236-0517} & 22 26 15.7 & $-$5 02 06 & 1.866 & 1103.4 & 5.08 & 2568.2 & 2620.6 & 0.7087 \\
 \object{Mrk 926} & 23 04 43.4 & $-$8 41 08 & 0.047 & 1101.0 & 3.51 & 2450.7 & 2470.3 & 0.6681 \\ \hline
 \object{PKS 0122-00} & 1 25 28.8 & $-$0 05 56 & 1.070 & 1155.4 & 3.29 & 1987.8 & 2193.8 & 0.5933 \\
 \object{PKS 0136+176} & 1 39 41.9 & 17 53 07 & 2.716 & 1157.2 & 5.02 & 2270.7 & 2491.6 & 0.6738 \\
 \object{PHL 1093} & 1 39 57.2 & 1 31 47 & 0.258 & 1156.4 & 3.24 & 2138.0 & 2353.9 & 0.6366\\
 \object{3C 208.0} & 8 53 08.6 & 13 52 54 & 1.109 & 909.6 & 3.56 & 2150.4 & 2351.4 & 0.6359 \\
 \object{3C 212.0} & 8 58 41.4 & 14 09 44 & 1.043 & 721.3 & 4.09 & 2328.1 & 2491.5 & 0.6738 \\
 \object{3C 216.0} & 9 09 33.5 & 42 53 45 & 0.668 & 539.0 & 1.40 & 2782.2 & 2996.6 & 0.8104 \\
 \object{4C 54.18} & 9 10 11.1 & 54 27 22 & 0.625 & 871.2 & 1.81 & 1899.5 & 2032.5 & 0.5497 \\
 \object{B2 0937+39} & 9 41 04.0 & 38 53 50 & 0.618 & 712.2 & 1.59 & 2133.0 & 2397.7 & 0.6484 \\
 \object{3C 254.0} & 11 14 38.5 & 40 37 20 & 0.734 & 1098.3 & 1.97 & 2317.0 & 2504.2 & 0.6772 \\
 \object{3C 270.1} & 12 20 33.9 & 33 43 12 & 1.519 & 1102.9 & 1.14 & 1989.8 & 2210.5 & 0.5978 \\
 \object{PKS 1351-018} & 13 54 06.8 & $-$2 06 03 & 3.709 & 1151.0 & 3.25 & 3058.7 & 3113.7 & 0.8421 \\
 \object{GC 1556+33} & 15 58 55.1 & 33 23 18 & 1.646 & 423.1 & 2.44 & 2962.3 & 3324.2 & 0.8990 \\
 \object{PG 1718+481} & 17 19 38.0 & 48 04 13 & 1.083 & 998.2 & 2.13 & 2249.4 & 2468.3 & 0.6675 \\
 \object{3C 446} & 22 25 47.1 & $-$4 57 01 & 1.404 & 1103.4 & 5.08 & 2651.5 & 2693.1 & 0.7283 \\
 \object{3C 454.3} & 22 53 57.6 & 16 08 53 & 0.859 & 943.2 & 7.06 & 3006.8 & 2935.9 & 0.7940 \\
\end{tabular*} 
\vspace{2mm}

\noindent
(a) ROSAT day, day 0 = 01 June 1990 \\
(b) Corrected total background counts, 10 ksec \\
(c) photons s$^{-1}$ deg$^{-2}$ \\

\end{table*}

As previously mentioned, these fields were imaged during \emph{ROSAT} PSPC
pointed observations.  In such observations, structures in the image caused by
the filter support struts are visible in each image.  In order to remove these
structures from the analysis, we selected only those photon events located
within a $20\arcmin$ radius of the detector center.

\subsection{The Scaling Index Method}

Suppose we have observed $N$ photons $(x_{1},\dots,x_{N})$ within a 
PSPC image. This map is described by the set $\{d_{ij}\}$ of 
distances between all photons. For each of the $N$ photons the 
cumulative number function is calculated
\begin{equation}
N_i(r)=\#\left\{j\big | d_{ij} \leq r\right\},
\end{equation}
where $\#\{j\}$ means the total number of elements in the set $\{j\}$.
We approximate the function $N_i(r)$ for each $i$ in some given range
$\lbrack r_1, r_2\rbrack$ with a power law
$N_i(r)\sim r^{\alpha_i}$ ($r_1 < r < r_2$)
and call the $\alpha_i$ ``scaling indices'' (or ``crowding indices'' 
as introduced by Grassberger \cite{grbapo}). 
Explicitly, the $\alpha_i$ are given by
\begin{equation}
\alpha_i = {\log {N_i(r_1)} - \log {N_i(r_2)}\over %
              (\log {r_1} - \log {r_2})}.
\end{equation}
The possible values of the scaling indices for a given scaling range
$\lbrack r_1, r_2\rbrack$ are determined by the conditional 
probability to find $n_2$ objects in a ball of radius $r_2$, if
a ball with the smaller radius $r_1$ contains $n_1$ objects. 
For a given scaling range $\lbrack r_1, r_2\rbrack$ only
those values of $\alpha_i$ will be observed for which two natural
numbers $n_1,n_2$ exist, such that
\begin{equation}
\alpha_i = {\log {n_1} - \log {n_2}\over %
            2 (\log {r_1} - \log {r_2})}
\end{equation}
and
\begin{equation}
Prob(N_i(r_2)=n_2 | N_i(r_1)=n_1) > 0.
\label{condprob}
\end{equation}
The latter conditional probability depends on the process 
that produced the observed photon map. 

Calculating for all photons $x_i$ the corresponding indices $\alpha_i$, 
we get the relative frequency distribution of scaling indices, 
or scaling index spectrum,
\begin{equation}
N_{freq}(\alpha)=%
\#\left\{\alpha_i | \alpha < \alpha_i < \alpha + %
\delta\alpha\right\}.
\end{equation}

Depending on the random processes in the considered field, 
$N_{freq}(\alpha)$ has a well-defined envelope and shows gaps,
where the probability of equation (\ref{condprob}) is zero. 

The search for sources or density variations in an otherwise 
homogeneous and isotropic background is synonymous
to the measurement of deviations from the expected frequency 
distribution. Any inhomogeneity or anisotropy will result
either in more power of the frequency distribution at
low $\alpha$-values or in filling up the discrete gaps. 
Since we have, in general, no precise knowledge about the
process producing the background and, as far as we know, no 
closed analytical description for $N_{freq}(\alpha)$ exists, 
another procedure is necessary to separate background 
photons from source photons.
 
For different scaling ranges $\lbrack r_1, r_2\rbrack$ 
the scaling indices are binned with $\delta\alpha=10^{-6}$, the computer 
precision in this case. The $r_1$ are chosen to be about the size of the 
expected sources, while $r_2$ $\approx 10 r_1$; in
our examples we typically used four different scaling ranges.
In a first step, the $\alpha$-values with $N_{freq}(\alpha)\leq 2$
are singled out for each scaling range.
In a second step, only those $\alpha$-values that have 
$N_{freq}(\alpha)\leq 2$ in at least two scaling ranges are considered
as belonging to source photons. Simulations showed that the
number of photons singled out with this procedure in a random
Poissonian field are less than $3\%$ of the total number. 

These spurious sources, caused by local fluctuations of the background, can be 
eliminated by means of a minimal spanning tree (MST) algorithm 
(Kruskal \cite{kruska}, Prim \cite{prim57}), accepting only those sources 
containing more than a set minimum number of photons. Since we want
to find close associations of photons, only those trees of the
MST are considered for which all edge-lengths are smaller than
a maximally allowed inter-photon separation $d_{max}$. The 
value of $d_{max}$ is determined by the mean inter-point 
separation expected in an isotropic and homogeneous point 
distribution. This separation is given by $d_{max}\approx\sqrt{A/N}$,
where $N$ is the total number of points and $A$ is the area of
the region where the $N$ points are found (see WSV \cite{wiedenmann}).

\begin{figure}
\resizebox{\hsize}{7.4cm}{\includegraphics{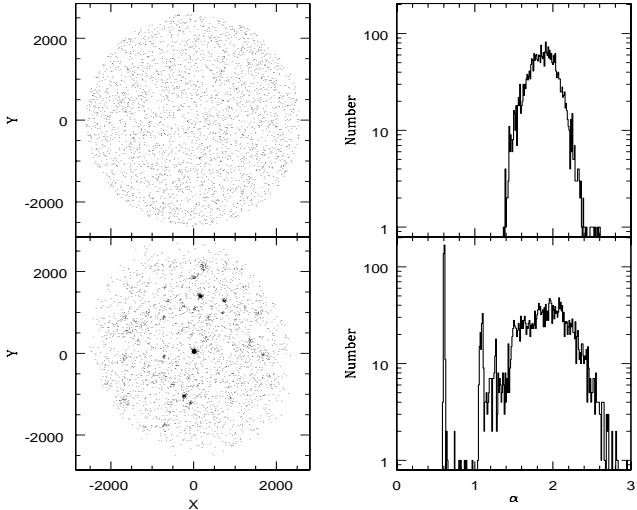}}
\caption{Photon maps and binned scaling indices $\alpha$ for a 
randomly-generated, smooth x-ray background (top) and an actual PSPC
image of 3C 216 (bottom). The control field contains 3000 randomly-placed 
photons, a typical background level for the brightest backgrounds in 
our sample and essentially equivalent to the 2997 photon background of 3C 216. 
The presence of sources broadens the $\alpha$
distribution, while brighter sources appear as sharp peaks at low
scaling index values.}
\label{fig:simspec}
\end{figure}

An example of the scaling-index spectrum is shown in Fig.~\ref{fig:simspec}.
The upper panels show a randomly-generated, smooth PSPC photon map (left)
and the scaling index $\alpha$-spectrum (right). In an infinite, uniform
image, the spectrum would be a delta function located at $\alpha =2$.
Because the field is finite and not infinitely homogeneous, the spectrum is
broadened and slightly offset.  The lower panels show an actual PSPC
image and its $\alpha$-spectrum. The spikes at low $\alpha$ are the
brightest sources, and the peak is broadened further due to the presence
of sources.

Analyses of randomly-generated smooth PSPC fields with background levels 
similar to those in our source fields revealed that the SIM detected
spurious sources with source counts up to $\sim 5$ photons. In 
order to ensure that no spurious sources were included in the source lists,
we set the lower limit on source counts to ten photons. This limit
undoubtedly excluded real sources, but as of yet there is no estimate
of source significance available for the SIM.

We applied the SIM algorithm to the fields in this sample, thereby creating
source lists and labeling each photon as a source or background photon.
Upon removal of the sources, it was obvious that the SIM had not correctly
identified all photons, as bright sources remained in the images. Further
investigation revealed that the SIM algorithm had difficulty identifying all
photons associated with a source if the photon density was very high.
We corrected this by identifying all photons within the SIM-defined source 
boundaries as source photons. We then applied a background correction based 
on the number of background photons one would expect to find under the source. 
While this method is less than ideal, the resulting background photon maps 
appear to have had all detected sources completely removed. 

In order to check the effectiveness of the scaling-index method, we also
determined the diffuse x-ray background flux of each field using the
maximum-likelihood source-detection algorithm included in the \emph{EXSAS}
data reduction software package (Zimmermann et al. \cite{zman}).
Our lower likelihood limit was set at 15, corresponding to a significance
level of $\sim 5\sigma$. We set the extraction radius at 2.5 times the FWHM
of each source. The background-corrected photon counts were then subtracted from 
the total photon count to give the raw background photon count. Vignetting 
corrections were not applied, since we assume that vignetting effects will
be similar in each image and since vignetting corrections were not available
for the SIM algorithm.

We find that the SIM analysis results in background levels that are higher
than the maximum-likelihood technique backgrounds by an average of
$158\pm 95$ counts.  The reason for this discrepancy is not fully clear and 
should be understood before the SIM is widely implemented. 
However, part of the discrepancy can be explained by noting that our SIM
source flux cut-off of 10 counts, while safely ignoring spurious sources,
almost certainly considered numerous true low-flux sources as spurious, 
resulting in these source photons being treated as background counts.

In this paper, we perform our analysis using the background count levels as 
calculated by the SIM. A reanalysis of the data using the numbers from the
maximum-likelihood technique resulted in the same qualitative conclusions,
although the resulting quantities do vary. The
SIM analysis has the benefit of identifying each background photon, which made
spectral analysis of the background photons straightforward.

\subsection{Correction for absorption by neutral hydrogen}
\smallskip

Despite our efforts to reduce the effects of absorption by neutral hydrogen, 
this absorption still affects the observed background photon counts measurably.
The fields in our sample cover a wide range of galactic $N_H$ values, from 
$1.13\times10^{20}$ cm$^{-2}$ to $7.06\times10^{20}$ cm$^{-2}$.  We corrected
the observed photon counts to a column density of zero. Correction 
factors were obtained using the most recent \emph{ROSAT} and \emph{ASCA} estimates of the 
cosmic x-ray background (CXRB) spectrum: a power-law with a photon index of 
1.42 and an 
intensity of 10.0 keV s$^{-1}$ cm$^{-2}$ sr$^{-1}$ keV$^{-1}$ superimposed 
on a Raymond-Smith thermal plasma emitting at a temperature of 0.142 keV 
with an emission measure of 18.8 in XSPEC/EXSAS units per steradian 
(Miyaji et al.~\cite{miyaji}, hereafter Mi98).  
The lower-temperature thermal plasma component 
($T\approx57$ eV) produced by the local hot bubble did not affect our 
model spectra due to its negligible flux in the observed photon energy band. 
The model spectrum was then projected through a layer of neutral hydrogen 
absorption corresponding to the  $N_H$ value for each quasar, and the 
resulting fractional flux decrease due to absorption was added back into the
raw background photon count of each field to produce the corrected background
counts given in Tab.~\ref{tab:1}.  $N_H$ values were taken from 
absorption line studies when available (Lockman \& Savage \cite{lockman}; 
Murphey et al. \thinspace\cite{murphey}), otherwise they were calculated from the
published HI maps of Dickey \& Lockman (\cite{dickey}).

\subsection{Spectral Analysis}
\smallskip

Taking the background photons as identified by the SIM, we used the 
\emph{EXSAS} spectral analysis packages to create spectra of the x-ray 
background in each field. We first attempted to fit the resulting spectrum to 
the spectrum of Mi98 described above. The goodness-of-fit was checked using a $
\chi^2$ test. If the reduced  $\chi^2$ indicated a poor fit, we calculated a
best-fit spectrum consisting of variable thermal and power-law components. 

\begin{figure}[t]
\resizebox{\hsize}{!}{\includegraphics{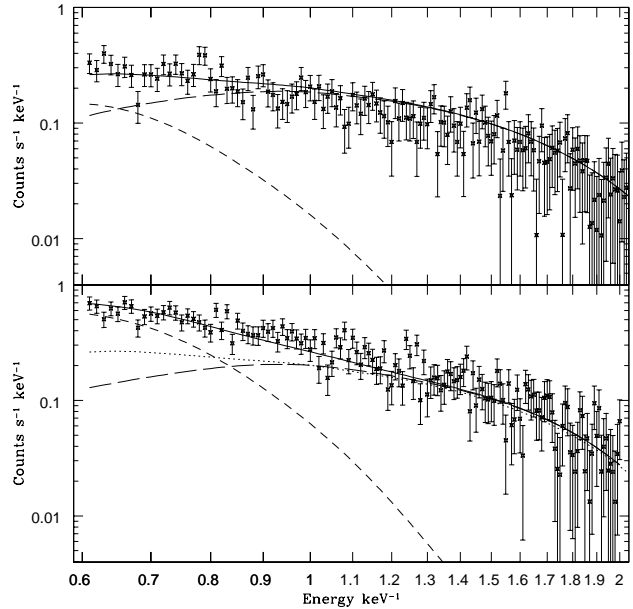}}
\caption{\emph{ROSAT} PSPC spectra of the SIM-selected background photons
for two radio-loud quasars -- \object{PKS 0136+176} (top) and 
\object{GC 1556+33} (bottom). The
top spectrum fits the theoretical diffuse spectrum of Mi98 (solid
line), but the same theoretical spectrum (dotted line) does
not fit \object{GC 1556+33}. The best fitting spectrum (solid line) is 
the Mi98 spectrum with an increased emission measure of the 0.142 keV 
component. In both panels, the short-dashed line indicates the
0.142 keV component, and the long-dashed line the power-law component of
the best-fit model. }
\label{fig:spec}
\end{figure}

Example spectra are plotted in Fig.~\ref{fig:spec}. The top spectrum, of
the diffuse x-ray background surrounding the radio-loud quasar 
\object{PKS 0136+176}, is fit well by the Mi98 background spectrum (solid 
line). The lower spectrum is of the diffuse x-ray background surrounding 
\object{GC 1556+33}. The Mi98 spectrum, shown as
a dashed line, clearly does not fit the data. The best-fitting spectrum,
shown as a solid line, resulted in the same power-law component and thermal 
plasma temperature as Mi98, only the emission measure of the 0.142 keV thermal
plasma was increased.  In fact, in virtually every background spectrum
a reasonable fit was achieved by varying this emission measure.  The 
best-fit emission measures varied from $15.9 \pm 1.7$ XSPEC units per
steradian for \object{4C 54.18} to $58.5 \pm 2.4$ XSPEC units per steradian
for \object{GC 1556+33} (Recall the Mi98 spectrum had an emission measure
of 18.8 XSPEC units per steradian). This suggests 
that the variations in diffuse background levels among the fields are due to 
variations in the hot thermal component of the CXRB, presumed by many to 
originate in the galactic halo (e.g. Gendreau et al.~\cite{gendreau}, Nousek
et al.~\cite{nousek}). This also suggests that
all variations in the diffuse backgrounds for the fields we examined cannot be
explained, as BSH propose, by emission from clusters of galaxies,
whose x-ray temperatures typically lie around a few keV rather than the 
0.142 keV modeled here (see, for example, David et al. \cite{david}).

The power-law component to the spectrum fits all of the observed fields rather
well. This agrees with the conclusions of Ishisaki (\cite{ishisaki}), who
found that, within systematic errors, the higher-energy diffuse x-ray 
background is flat on angular scales of $\sim 1\degr$

\section{Discussion}

\begin{figure}[t]
\resizebox{\hsize}{!}{\includegraphics{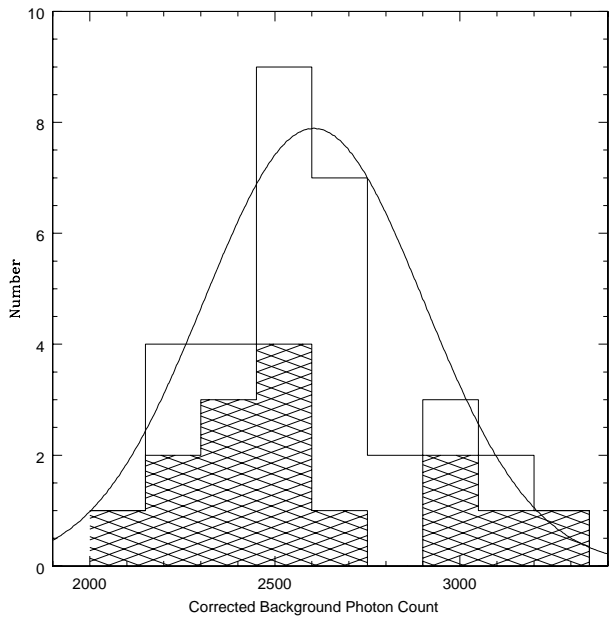}}
\caption{A histogram of the total observed background counts for the quasar
sample.  The counts are corrected for $N_H$ and represent the photons 
collected during a 10\,050 second exposure covering $\sim0.42$ deg$^2$ of
sky.  The white region indicates the data for all quasars, and the hatched
region indicates the data for radio-loud quasars only. The solid line 
indicates the Gaussian fit to the entire sample.}
\label{fig:ncts}
\end{figure}

\subsection{Diffuse x-ray background level distribution}

The distribution of diffuse x-ray background counts for our sample is shown in
Fig.~\ref{fig:ncts}.  The entire sample has a mean of $2606.0 \pm 296.6$
counts for the 10 ksec exposure, corresponding to a mean detected flux of
0.7048 photons s$^{-1}$ deg$^{-2}$. The distribution can be modeled by a
normal distribution, shown in the figure as a solid line. A simple
Kolmogorov-Smirnov (K-S) test results in a significance level of 0.957, 
indicating a fairly robust fit.

Examination of the radio-loud and radio-quiet quasar fields separately 
brings one
to slightly different conclusions.  The backgrounds surrounding radio-quiet 
quasars have a mean of $2635.5\pm 226.7$ integrated counts, corresponding
to a mean detected photon flux of 0.7127 photons s$^{-1} \deg^{-2}$. This is
slightly higher, but not significantly so, than the mean background flux
for the entire quasar sample.  A K-S test comparing the radio-quiet
quasars with a Gaussian distribution of the same mean and standard deviation
gives a significance level of 0.953, which again indicates a rather robust fit.

The diffuse background surrounding radio-loud quasars has more ambiguous
results. The radio-loud quasar background distribution is shown in 
Fig.~\ref{fig:ncts} as the hatched histogram.  At first glance, the
distribution gives an indication of being bi-modal, but the
small-number statistics involved means that such qualitative observations
can be misleading.  The diffuse x-ray backgrounds in the radio-loud quasar
fields had a mean of $2570.6 \pm 369.0$ detected counts,
corresponding to a flux of 0.6952 photons s$^{-1} \deg^{-2}$, which is 
slightly lower, but not significantly so, than the overall mean background
population. If we try to model the radio-loud background distribution by
a Gaussian, the K-S significance level is 0.363, a level not ruling out a
Gaussian fit, but not strongly supporting such a fit.  Moreover, a K-S test
comparing the radio-loud and radio-quiet background distributions gives only
a 0.168 likelihood that the two arise from the same distribution. Again,
this does not rule out that the two arise from the same distribution, but it
does not support such a claim, either.

\subsection{Arguments against clusters of galaxies}

Due to the ambiguity of the results concerning the diffuse x-ray background
surrounding radio-loud quasars, we searched for explanations supporting
a two-population hypothesis. Because of the spectral analysis
noted before, we consider it highly unlikely that these enhancements
are due to low-luminosity x-ray clusters of galaxies as BSH
suggest.  Further evidence rejecting the galaxy-cluster explanation came
from an analysis of the smoothness of each field once sources were removed.
The remaining fluctuations in the x-ray background were not statistically 
significant, implying that any x-ray-emitting galaxy clusters would have to
be $\gtrsim 20\arcmin$ in angular size. One would expect that galaxies in 
clusters of this size would be visible at optical wavelengths. A search of 
the $20\arcmin$ region surrounding
each quasar in the NASA/IPAC Extragalactic Database (NED) revealed no known 
galaxy clusters which were not detected as 
sources and removed by the scaling-index method.

\subsection{Galactic origins of enhancements}

Given the apparent demise of the galaxy cluster hypothesis for the origin of
the diffuse x-ray background enhancements, we need to develop
another hypothesis. One likely source of large-scale x-ray emission is an
extended galactic object.  We located each field in the $\frac{3}{4}$ keV-band 
high-resolution \emph{ROSAT All-Sky Survey} maps of Snowden et al. 
(\cite{snowden2}).  Two of the radio-loud quasar fields with enhanced
background emission, \object{GC 1556+33} and \object{PKS 1351-018}, lie in 
regions of slightly enhanced x-ray emission due to Loop I. Also, the field 
surrounding \object{3C 216} lies in a small region of larger-scale enhanced 
emission. These enhancements were not obvious on
the low-resolution RASS maps (Snowden et al. \cite{snowden})
used in the sample selection. Therefore, it is reasonable to assert that the
excess emission in these three fields is due to variations in galactic
emission. The other radio-loud quasar field showing enhanced emission, 
\object{3C 454.3}, does not appear to lie in regions of large-scale diffuse 
x-ray enhancements on the high-resolution RASS maps.

We also inspected several infrared images of the sample fields taken from
the Infrared Astronomical Satellite (IRAS) data archives. A brief comparison
showed no obvious connection between the diffuse x-ray background levels and
infrared emission. This could suggest that our corrections for x-ray absorption
were sufficient, though it also could suggest that the observed background
enhancements are not galactic in origin. The data here is too sketchy to
permit a conclusion.

\subsection{Possible systematic origins of enhanced background levels}

\begin{figure*}[t]
\resizebox{\hsize}{16 cm}{\includegraphics{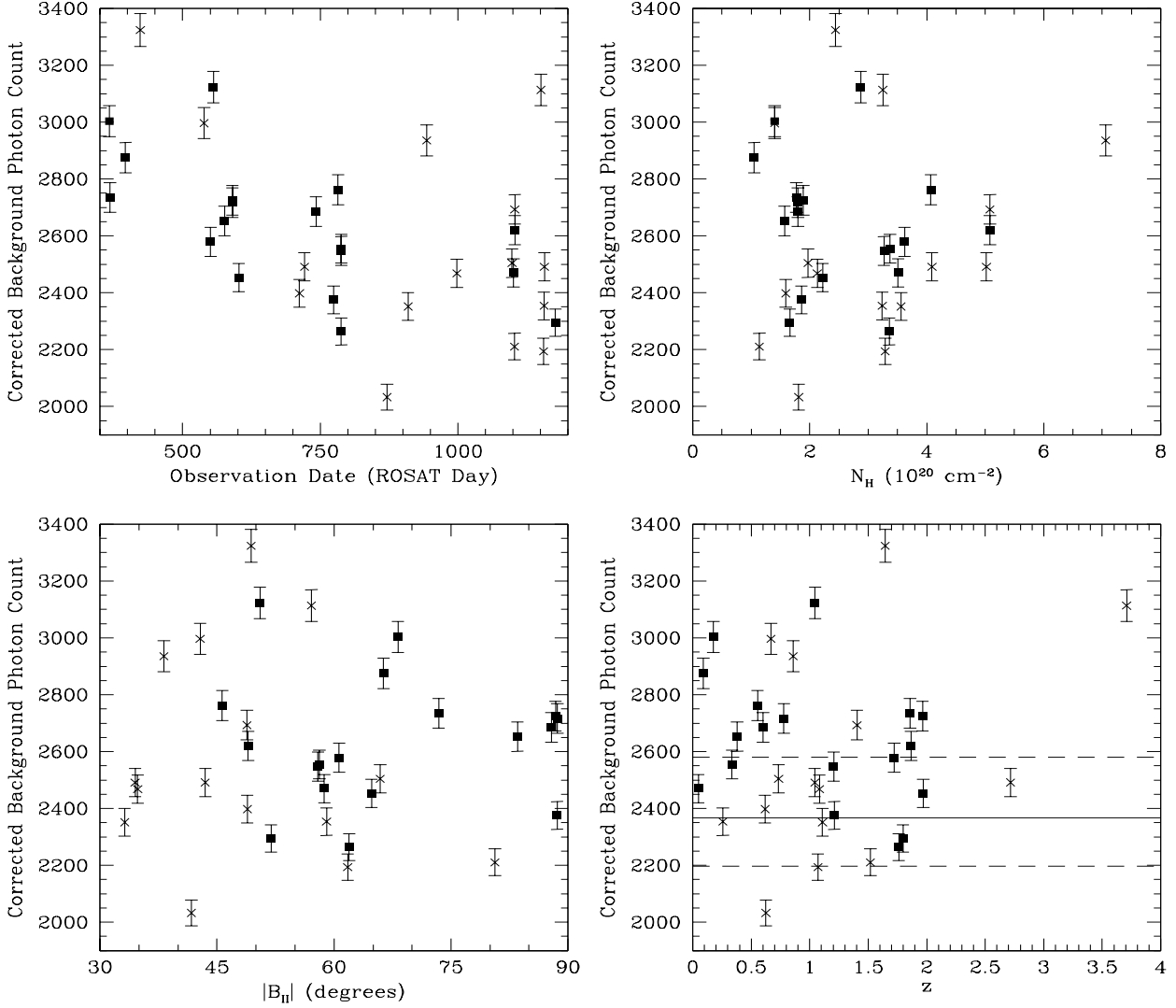}}
\caption{Plots showing the total corrected background counts as functions of
the times of observation, the galactic $N_H$ values, the absolute value
of the galactic latitude, and redshift.  Filed squares represent radio-quiet
quasars; crosses represent radio-loud quasars. In the plot of background counts
versus redshift, the CXRB spectra as determined by Miyaji
et al. (\cite{miyaji}) and folded through the ROSAT PSPC response function is
plotted.  The solid line indicates best-fit values, and the dotted lines 
indicate the 90\% errors quoted by Mi98. The error bars shown are
$1\sigma$ error bars assuming Poisson photon counting statistics. No 
systematic errors are included.}
\label{fig:bkgds}
\end{figure*}

We have searched for other systematics which could explain the
variations in diffuse x-ray background levels.  Fig.~\ref{fig:bkgds} shows
the detected background counts as compared with observation date, 
galactic neutral hydrogen column densities, galactic latitudes, and quasar
redshifts. The plots show no discernible correlation of background
counts with these quantities, with the possible exception of observation
date. 

The ROSAT PSPC has been known to show variations in response with time
(Prieto, Hasinger, \& Snowden 1996), but these effects have been accounted
for in the data reduction. A least-squares linear regression fit results
in a $2\sigma$ detection of a downward trend. Such a decrease of sensitivity,
on the order of 10\%, would have important implications for all PSPC data
reduction.  The fact that no such decrement has been noticed by others casts
doubt onto this hypothesis. An analysis of a much larger sample of
fields, or of the same field at various epochs, would be needed to prove
that a change of sensitivity with time in fact occured.

Also shown in Fig.~\ref{fig:bkgds} is the cosmic x-ray background spectrum
of Mi98, integrated over $10\,050$ ksec and folded through the ROSAT PSPC
response function (solid line) and the 90\% confidence limits (dashed line). 
This spectrum appears to underestimate the observed 
background. Recalling that the SIM analysis tended to give
backgrounds nearly 200 counts higher than other analysis techniques brings 
the Mi98 spectrum into reasonable agreement with the
observed background levels.

\begin{figure}[t]
\resizebox{\hsize}{!}{\includegraphics{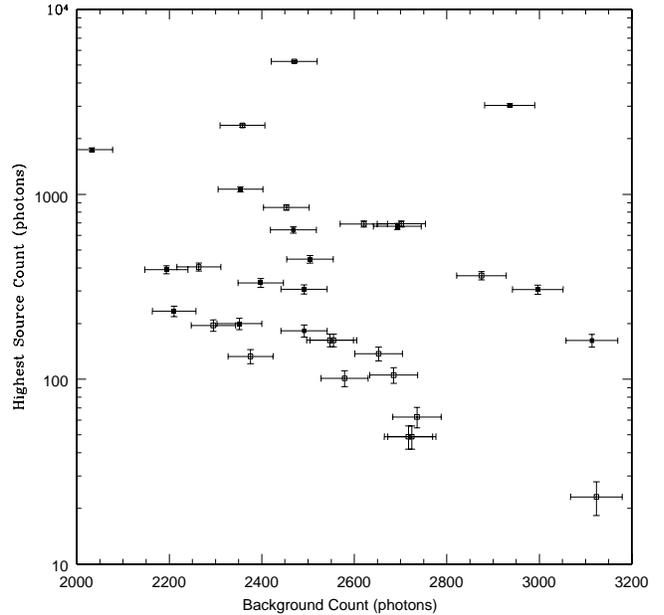}}
\caption{The relationship between the detected counts of the brightest source
in each frame is compared with the background levels. Although a Spearman
rank-order correlation test gives a weak detection of a correlation 
(2.29$\sigma$), the large scatter (note the logarithmic scale on the vertical
axis) and considerations discussed in the text cast some doubt
on this detection. Error bars are the 1$\sigma$ Poisson noise only. Solid
blocks are for fields containing radio-quiet quasars, while open blocks
indicate the fields containing radio-loud quasars. }
\label{fig:fig5}
\end{figure}

Given the discrepancies between the SIM method and more traditional 
source-detection routines, it is worthwhile to examine possible correlations
between source fluxes and the background levels. A Spearman rank-order
correlation test revealed no significant correlations between quasar flux and
the background flux or between total flux of detected sources and the
background flux. The same test gave a 2.29$\sigma$-likelihood of a
correlation between each field's  brightest-source flux and the background
flux. The plot of this correlation is given in Fig.~\ref{fig:fig5}, and 
suggests that the fields with the brightest backgrounds have dimmer main
sources. However, this perception is heavily biased by the lowest-flux source.
Additionally, one would expect that, if the SIM were not adequately removing
source photons, the brighter sources should leave a higher number of photons
in the background, the opposite from what is seen. Still, this possible
correlation should be kept in mind.

Despite the ambiguous statistical results, the above analysis and
the spectral analysis of the variations in the diffuse x-ray background
surrounding quasars strongly suggests that these variations are fully
explainable by variations in the hot, galactic (i.e. 0.142 keV) component of 
the diffuse x-ray background. The analysis also shows that there is no 
statistically significant difference between the diffuse x-ray backgrounds 
in the regions surrounding radio-loud and radio-quiet quasars.

\section{Conclusions}

We have examined the large-scale ($\lesssim 20\arcmin$) diffuse x-ray 
background
surrounding 15 radio-loud and 18 radio-quiet quasars. To accomplish this, we
have used a new source-detection algorithm, the scaling-index method. This
method is not yet fully refined, with three major uncertainties remaining. The
first of these is the lowest flux at which actual and spurious sources can
be discriminated. A second concern is that the background fields so examined
had higher detected x-ray emission levels than the same fields examined
by other source-detection algorithms. Our final concern is that the SIM has
difficulty identifying source photons in bright sources, where the photon
density is highest. The SIM does offer the distinct advantage of
separating background and source photons, as well as the potential to 
locate the true extent of extended sources.

Using the SIM, we find that the backgrounds can be adequately examined by
a single Gaussian distribution with a mean of 2606.0$\pm296.6$ detected 
photons in a 10 ksec exposure. Spectral analysis of these backgrounds
suggests that the variations in the background levels is due to fluctuations
in the previously-reported 0.142 keV component of the cosmic x-ray background.
There is some evidence that there may be a separate population of enhanced 
x-ray  backgrounds surrounding some radio-loud quasars, but there is no
plausible source or systematic explaining such a division. There is
no evidence of large-scale diffuse emission from x-ray clusters being
responsible for notable enhancement of these backgrounds, as had been 
hypothesized by previous analyses. Finally, we find no 
statistically-significant, systematic differences between the large-scale 
diffuse x-ray 
backgrounds surrounding radio-loud quasars and those surrounding
radio-quiet quasars. This suggests that x-ray evidence for differences in the
local environments of quasars is either present only on angular scales of
less than a few arcminutes, present only at very low flux levels, or 
non-existent.

\noindent
\begin{acknowledgements}
The authors wish to express appreciation to M. Freyberg for helpful discussions
concerning data analysis, especially in the realm of the diffuse x-ray
background.  Appreciation is also expressed to H. Scheingraber for discussions 
on the SIM. The authors also wish to thank the anonymous
referee for helpful suggestions in improving this paper.
KAW acknowledges the financial support of a Fulbright Fellowship from the 
Fulbright-Kommission, Bonn, Federal Republic of Germany and the hospitality
of the MPE during his Fulbright tenure.  KAW would also like to thank
G.R. Penn and K. M\"uller-Osten for their kindness, support, and skiing 
lessons.  This research has 
made use of the NASA/IPAC Extragalactic Database (NED) which is operated by 
the Jet Propulsion Laboratory, California Institute of Technology, under 
contract with the National Aeronautics and Space Administration. 
\end{acknowledgements}


\begin{thebibliography}{}
\bibitem[1994]{bartelmann} 
	Bartelmann, M., Schneider, P., Hasinger, G., 1994, A\&A 290, 399 (BSH)
\bibitem[1988]{boyle}
	Boyle, B.J., Shanks, T., Yee, H.K.C., 1988, in {\it Large Scale
	Structures of the Universe, IAU Symposium No. 130}, 
	ed. J. Audouze, M.-C. Pelletan, and A. Szalay, p. 576
\bibitem[1993]{briel}
	Briel, U.G., Henry, J.P., 1993, A\&A 278, 379
\bibitem[1994]{burg}
	Burg, R., Giacconi, R., Forman, W., Jones, C., 1994, ApJ 442, 37
\bibitem[1993]{david}
	David, L.P., Slyz, A., Jones, C., et al., 1993,
	ApJ 412, 479
\bibitem[1990]{dickey}
	Dickey, J.M., Lockman, F.J., 1990, ARA\&A 28, 215
\bibitem[1997]{fried}
	Fried, J.W., 1997, A\&A 319, 365
\bibitem[1995]{gendreau}
	Gendreau, K.C., Mushotozky, R., Fabian, A.C., et al., 1995, PASJ 47,
	L5
\bibitem[1988]{grbapo}
	Grassberger, P., Badii, R., Politi, A., 1988, J. Stat. Phys. 51, 135
\bibitem[1995]{hall}
	Hall, P.B., Ellingson, E., Green, R.F., Yee, H.K.C., 1995, AJ 110, 
	513
\bibitem[1997]{ishisaki}
	Ishisaki, Y., 1997, Ph.D. Thesis, University of Tokyo
\bibitem[1956]{kruska}
	Kruskal, J.B., 1956, Proc. Am. Math. Soc. 7, 48
\bibitem[1995]{lockman}
	Lockman, F.J., Savage, B.D., 1995, ApJS 97, 1
\bibitem[1998]{mchardy}
	McHardy, I., Jones, L., Merrifield, M., et al., 1998, MNRAS 295, 641
\bibitem[1998]{miyaji}
	Miyaji, T., Ishisaki, Y., Ogasaka, Y., et al., 1998, A\&A 334, L13
	(Mi98)
\bibitem[1996]{murphey}
	Murphey, E.M., Lockman, F.J., Laor, A., Elvis, M., 1996, ApJS 105,
	369
\bibitem[1982]{nousek}
	Nousek, J.A., Fried, P.M., Sanders, W.T., Kraushaar, W.L., 1982,
	ApJ 258, 83
\bibitem[1980]{pfeffer}
	Pfeffermann E., Briel U.G., Hippmann H., et al., 1986, Proc. SPIE 733,
	519
\bibitem[1996]{prieto}
	Prieto, M.A., Hasinger, G., Snowden, S.L., 1996, A\&AS 120, 187
\bibitem[1957]{prim57}
	Prim, R.C., 1957, Bell Syst. Tech. J. 36, 1389
\bibitem[1995]{snowden}
	Snowden, S.L., Freyberg, M.J., Plucinsky, P.P., et al., 1995, 
	ApJ 454, 643
\bibitem[1997]{snowden2}
	Snowden, S.L., Egger, R., Freyberg, M.J., et al., 1997, ApJ 485, 125
\bibitem[1993]{veron}
	V\'eron-Cetty, M.P., V\'eron, P., 1993, ESO Scientific Report 13
\bibitem[1996]{voges}
	Voges, W., Boller, Th., Dennerl, K., et al., 1996, in {\it 
	R\"ontgenstrahlung from the Universe}, ed. H.U. Zimmermann, J. 
	Tr\"umper, and H. Yorke, MPE Report 263, p.637
\bibitem[1997]{wiedenmann}
	Wiedenmann, G., Scheingraber, H., Voges, W., 1997, in {\it Data
	Analysis in Astronomy}, ed. V Di Ges\'u, M.J.B. Duff, A. Heck,
	et al. (Singapore: World Scientific) (WSV)
\bibitem[1992]{zman}
	Zimmermann, H.U., Belloni, T., Boese, G., et al., 1992, in {\it Data
	Analysis in Astronomy IV}, ed. V. Di Gesu, L. Scarsi, R. Buccheri, 
	et al., p.141 (New York: Plenum Press)

\end{thebibliography}
\end{document}